**Detectability Simulations of a NIR Surface Biosignature on Proxima Centauri b with Future Space Observatories**

Short title: Proxima b Surface Biosignature Detectability

*Connor O. Metz*[*,1], *Nancy Y. Kiang*[2], *Geronimo L. Villanueva*[3], *Mary N. Parenteau*[4], *Vincent Kofman*[3,5]

[*]*Corresponding author: cometz@umich.edu*
[1]*Department of Climate and Space Sciences and Engineering, University of Michigan, 2455 Hayward St, Ann Arbor MI 48109*
[2]*NASA Goddard Institute for Space Studies, Code 611, 2880 Broadway, New York NY 10025*
[3]*NASA Goddard Space Flight Center, Code 690, 8800 Greenbelt Rd., Greenbelt, MD 20771*
[4]*NASA Ames Research Center, Moffett Field, CA 94035*
[5]*American University, 4400 Massachusetts Ave NW, Washington, DC 20016*

**Abstract**
  Telescope missions are currently being designed which will make direct imaging of habitable exoplanets possible in the near future, and studies are needed to quantify the detectability of biosignature features in the planet's reflectance spectrum. We simulated the detectability of a NIR-absorbing surface biosignature feature with simulated observations of the nearby exoplanet Proxima Centauri b. We modeled a biosignature spectral feature with a reflectance spectrum based on an anoxygenic photosynthetic bacterial species that has strong absorption at 1 um, which could make it well suited for life on an M-dwarf hosted planet. We modeled the distribution of this organism across the planet's surface based on climate states from a 3D General Circulation Model (GCM), which were Archean and Proterozoic-like exo-Earth analogues. We included the GCM runs' prognostically simulated water clouds and added organic haze into the Archean-like atmospheres. We simulated observations of these Proxima Centauri b scenarios with the LUVOIR-A and B telescope concepts, with LUVOIR-B serving as a proxy to the planned Habitable Worlds Observatory (HWO). We calculated integration times necessary to detect the biosignature, and found that it would be detectable on Proxima Centauri b if the organism is moderately abundant (greater than a 1-4% global surface area coverage), as long as the atmosphere is transmitting in the wavelength range under consideration. Small amounts of methane, clouds, and haze do not greatly impede detectability. We found preliminary evidence that such a biosignature would be detectable on exoplanets within 15 pc, but further investigations are needed to corroborate this.

## 1. Introduction

  The search for life elsewhere in the universe is an ancient question, but still a relatively young discipline of science. Identifying a sign of life on another planet (known as a "biosignature") is an extremely difficult task, for two reasons: (1) the necessary astronomical observations require very high levels of precision, and (2) because it is unclear what exactly we are looking for. Life on Earth is only one example of life, and the biochemistry on other planets may be quite different from what we are familiar with. The focus of the present work is item (1), the astronomical technique and technology required for such detections. Much of the work in that field to date has focused on atmospheric biosignatures (e.g., Checlair *et al.* 2021, Domagal-Goldman *et al.* 2011), primarily because exoplanet atmospheres are more easily detectable than surfaces with the current generation of instruments and techniques (for one, because of refraction when observed in the transit geometry, see García Muñoz *et al.* 2012). Some previous studies have simulated reflectance spectra of exoplanets with potential surface biosignatures (e.g, Schwieterman *et al.* 2015, Sanromá *et al.* 2014) but many of them have not simulated noise in the spectra, which makes it difficult to constrain the integration time required to detect these biosignatures.
  To observe a spectral biosignature, one needs two things: (1) a recognizable feature in the spectrum that is indicative of the presence of life, and (2) a sufficiently high signal to noise ratio (SNR) in that wavelength region of the spectrum. A typical requirement for item (2) is that the signal be at least 5 times higher than the standard deviation of the noise. Hereafter, satisfying item (1) will be referred to as "interpretability," and satisfying item (2) will be referred to as "detectability." This work aims to calculate integration times required to detect surface biosignatures by simulating spectral observations with future space observatories. It is essential to characterize integration times in advance of observations so that one can know whether or not it is possible to detect desired spectral signatures originating from the target with a given telescope. Telescope time is a rare commodity, and many thousands of hours would likely never be allocated to observing a single target.

Plans for future space telescopes are currently in development that will be capable of observing reflectance spectra from small terrestrial planets, which will enable the study of Earth-like exoplanetary surfaces and open the possibility of detecting surface biosignatures. Specifically, the Habitable Worlds Observatory (HWO) concept is currently in development by NASA, which builds upon the LUVOIR (Large UV/Optical/IR Surveyor) and HabEx (Habitable Exoplanet Observatory) concepts (Clery 2023, The LUVOIR Team 2019, Gaudi *et al.* 2020). At present, the instrument design specifications of HWO are under development and remain highly uncertain. Detection feasibility studies such as the present work are necessary for the development of future space telescopes, to see if detection of surface biosignatures would be an attainable goal for the mission (and what observation requirements would be necessary to make it attainable).

In the search for life on exoplanets, surface biosignatures provide an alternative to atmospheric signatures on planets that lack atmospheric biosignatures. It would also be of high scientific value to detect both an atmospheric and surface biosignature on a single planet. Detection of a surface biosignature would bolster confidence that an atmospheric signature (such as molecular oxygen) is actually biological in origin. A good deal of work has been done in the field about detectability and interpretability of Earth's most prominent surface biosignature, the vegetation red edge (VRE). Sagan *et al.* (1993) detected the VRE in spectra from Galileo's flyby of the Earth, and posited that it would be an interpretable biosignature, when considered in combination with widespread $H_2O$, $O_2$, and the strong $CH_4$–$O_2$ chemical disequilibrium in the atmosphere. Other studies have found the VRE detectable in the Earthshine (Seager *et al.* 2005; Ford *et al.* 2001; Woolf *et al.* 2002; Montañes-Rodriguez *et al.* 2006), and explored its interpretability on the Earth at different phases (Tinetti *et al.,* 2006a*), and* through paleo-time (Arnold *et al.* 2008a, Kaltenegger et al., 2007). Seager *et al.* (2005) argued that the spectral features of photosynthetic absorption should be common in putative extraterrestrial biology; thus the VRE detectability has been explored for exo-Earths (Gomez Barrientos *et al.* 2023).

Photosynthetic alternatives to the VRE have been proposed, although few studies have explored their detectability. Tinetti et al. (2006b) constructed an NIR-edge for M-star hosted planets, finding its noise-free spectral contrast (relative to continuum) could potentially exceed that of the red edge on Earth. Sanromá *et al.* (2014) characterized the interpretability of spectral signatures due to primitive purple anoxygenic phototrophs on Archean-like planets. DasSarma and Schwieterman (2018) examined a different type of 'purple Earth' and posited that Archaea, containing purple-colored bacteriorhodopsin, may have dominated the early Earth prior to the evolution of photosynthesis. Coelho *et al.* (2024) designed a database of purple bacteria reflection spectra, which can be used for future modeling studies such as the present work, with the idea that NIR absorption might be an evolutionary advantage on M-dwarf hosted planets. Other studies have examined biosignature potential of "non-photosynthetic" pigments (Schwieterman et al. 2015) and other biological molecules (Poch et al. 2017), and developed extensive databases of reflectance spectra of extremophilic microbes (Hegde et al. 2015) and microbes in icy environments (Coelho et al. 2022). Borges *et al.*(2024) simulated the detectability of full global cover of four different microbial mats on exoplanets, using 1D (globally averaged) models and plausible observational parameters for HWO. They measured detectability as a difference relative to similar mineral features, under various cloud cover conditions and placing the planet at various distances from Earth.

This study aims to constrain the detectability of a surface biosignature whose distribution is consistent with the climate for Archean and Proterozoic-like, for the limiting case of the nearest exoplanet, Proxima Centauri b (Anglada-Escudé *et al.* 2016). Proxima Centauri b, because of its proximity (at 1.3 pc away) and high absolute brightness, would be relatively favorable to observe compared to more distant exoplanets. The small star-planet angular separation makes coronagraphic observations of this system difficult, but they would theoretically be possible with LUVOIR A and B (The LUVOIR Team 2019). Just from distance considerations alone (not considering coronography star-planet angular separation), if surface biosignatures would be undetectable on Proxima Centauri b (hereafter referred to as "Proxima b"), then there would likely be little hope of detecting them on farther exoplanets, and vice versa (if they are detectable on Proxima b, there is hope of detection on more distant planets).

Proxima b has been the subject of many studies in exoplanetary science due to its possible habitability and proximity to Earth (Ribas *et al.* 2016, Turbet *et al.* 2016, Del Genio *et al.* 2019, etc.). Proxima b orbits its host star with a semi-major axis of 0.0485 AU and an orbital period 11.2 days, which places it in the habitable zone of its host star (an M5.5 dwarf, with an effective temperature of ~3050 K) (Anglada-Escudé *et al.* 2016). We choose to study a photosynthetic microbe with NIR-absorbing pigments, which can serve as surface biosignatures because Proxima Centauri outputs mostly NIR light. Such an organism could be well suited for life around an M-dwarf star (following the logic of Kiang *et al.* 2007). Proxima b was discovered with the radial velocity technique (Anglada-Escudé *et al.* 2016). Its composition remains unknown at the time of writing. Ribas *et al.* (2016) demonstrated that Proxima b could sustain liquid water on its surface, and several studies have attempted to constrain its potential habitability under different atmospheric and surface composition scenarios (e.g. Turbet *et al.* 2016, Del Genio *et al.* 2019, Meadows *et al.* 2018).

To best capture some of the complexity involved in detecting surface biosignatures, this study uses simulated climate states of Proxima b from a 3D General Circulation Model (GCM) in combination with an observation simulator to estimate the reflection spectrum of Proxima b as might be seen by future space telescopes. The advantage of using 3D models in surface biosignature detectability simulations is greater realism. 3D models can prognostically simulate clouds, capturing their heterogeneity. By contrast, in 1D climate models cloud effects must be prescribed. Because 3D models simulate the diversity in climate zones of a planet, they also can capture the complexities of distributions of life across the planetary surface. Even though upcoming exoplanet observing missions will only observe the planet in a disk or phase average, 3D atmospheric and surface heterogeneities still have important effects on the observed/simulated spectrum. Berdyugina and Kuhn (2019) theoretically demonstrated a technique using phase curves to indirectly image the surface of exoplanets, with applications to Proxima b. They showed that one could reconstruct a map of photosynthetic life on an exo-Earth, though doing so for Proxima b would require very long integration times, and rotational maps may be complicated for tidally locked planets as the planet enters the inner working angle (IWA).

## 2. Methods

### *2.1 Simulated Proxima b Scenarios*

The GCM outputs used in this study are taken from Del Genio *et al.* (2019), hereafter referred to as "DG19." DG19 modeled a wide variety of scenarios for Proxima b using the ROCKE-3D GCM (Resolving Orbital and Climate Keys of Earth and Extraterrestrial Environments with Dynamics) (Way *et al.* 2017). ROCKE-3D simulates atmosphere and ocean dynamics and radiative transfer with arbitrary atmospheric/oceanic composition and stellar spectral radiance. Like all GCMs, ROCKE-3D divides the planet (atmosphere, ocean, and surface) into grid cells and solves dynamical equations in discrete timesteps until a climatological equilibrium is reached. See Way *et al.* (2017) for an in-depth review of ROCKE-3D's methodology and capabilities. The DG19 GCM simulations did not explicitly include any biological activity. In the present work, spectral features of life are added to the runs in the observation simulator.

In this work, outputs from three of the GCM runs from DG19 are used to simulate observations of Proxima b. Table 1 contains descriptions of the atmospheric compositions, other basic parameters of the runs, and some of their results. The *Archean* runs' atmospheric compositions are based on what the Earth's atmosphere could have been like during the Archean eon, having large amounts of methane. *Control* is an anoxic Earth-like planet with moderate greenhouse gases, and so is used in the present work as a Proterozoic-like analogue. See DG19 for in-depth discussion of experimental setup and results. The planets in these three runs from DG19 are all aquaplanets with only water ice and liquid water on the surface, the spatial distribution of which is shown in Figure 3 in the column labeled $f_{ow}$. The coupled atmosphere and ocean, along with the heat transport of the ocean, creates unique open water/ice coverage patterns on each DG19 run. All DG19 runs contained water ice and liquid water clouds in their atmospheres. In the DG19 runs, water and aerosols vary in their abundance in each grid cell, whereas all other atmospheric species are held constant across the planet. DG19 averaged the climatological outputs over long time scales (~1000 orbits), which is suitable for this work which simulates long integration time spectral observations. The three GCM states used in this work were all tidally locked in a 1:1 spin-orbit resonance, which is likely to be the case for Proxima b, or it could be in a 3:2 spin-orbit resonance (Ribas *et al.* 2016), though Barnes *et al.* (2016) consider the 3:2 case unlikely.

**Table 1**

Summary of Scenarios from Del Genio *et al.* (2019) Used in the Present Work

| DG19 Scenario Name | Description | $H_2O$ at 0.1 $P_{surf}$ (ppmv) | Mean Surface T (ºC) | $f_{hab}$ | Relevant Notes |
|---|---|---|---|---|---|
| *Control* | 0.984 bar $N_2$ + 376 ppmv $CO_2$, dynamic ocean, aquaplanet, synchronous rotation | 1.1 | −21 | 0.42 | Mostly clear sky, Proterozoic analogue |
| *Archean Low* | Like *Control* but with 638 ppmv $CO_2$ and 450 ppmv $CH_4$ | 108.2 | −15 | 0.50 | More greenhouse gasses warm the climate, more open ocean, more humid |
| *Archean High* | Like *Control* but with 10,000 ppmv $CO_2$ and 2000 ppmv $CH_4$ | 439.8 | −10 | 0.56 | Very cloudy on eastern half of dayside |

Note: $f_{hab}$ indicates the global fraction of the planet's surface that is covered in liquid water. This table contains only partial information about the scenarios. For full details of simulation setup and results, see DG19 Tables 4 and 5. Note that looking only at a scenario's mean surface temperature only gives a rough idea of the climate.

It is interesting to simulate the detectability of surface biosignatures on Archean and Proterozoic-like Earth analogues, considering the "Earth as an exoplanet" biosignature search strategy (e.g., in the LUVOIR Final Report, The LUVOIR Team 2019). In short, this strategy assumes that other inhabited planets may follow similar evolutionary trajectories to that of the Earth, since Earth is the only inhabited planet we know of at present. Hypothetically, an exoEarth's evolutionary development should be based solely on internal planetary factors such as interior evolution and geologic processes such as volcanic activity, which release reductants to the atmosphere that provide evolutionary selective pressure for the development of certain metabolisms (e.g., Pierson and Olson, 1989).It should not necessarily depend on the absolute age of the system, so Archean-Earth analogues may be present in systems older than a few billion years. Although methane and/or organic haze could have been a detectable biosignature during the Archean period of Earth's history (Guzmán-Marmolejo *et al.* 2013, Schwieterman *et al.* 2018, The LUVOIR Team 2019), these atmospheric signatures are difficult to interpret (because sources of methane can be both biologic and abiologic) and thus may not be enough by themselves to confirm the presence of life on an exoplanet. At time of writing, no directly observable biosignature has been confirmed for Proterozoic-like exo-Earths (The LUVOIR Team 2019). It has been hypothesized that biogenic methane or nitrous oxide could have been present on Proterozoic Earth in large concentrations (and thus remotely detectable), but the concentrations of these gasses in the past are unknown. Similarly, the oxygen and ozone concentrations during the Proterozoic are unknown, but could have been remotely detectable if present in large quantities (Schwieterman *et al.* 2018, The LUVOIR Team 2019, Robinson and Reinhard 2020). Interestingly, an indirectly observable biosignature was recently proposed for Proterozoic-like planets by Young *et al.* (2024), who demonstrated the feasibility of using the $CH_4$-$O_2$ disequilibrium pair to infer the presence of life (supposing those gasses were generated during the Proterozoic in sufficient quantities). The present work aims to estimate how feasible it would be to directly detect surface biosignatures with current technology, which would augment our search for directly detectable signs of life on Archean and Proterozoic-like exo-Earths.

Observing a surface biosignature in conjunction with an atmospheric biosignature would significantly improve the confidence of an extrasolar life detection. However, the presence of methane also complicates the detection of the surface biosignature under investigation in this work because of the spectral overlap of methane absorbance with the surface biosignature feature. Additionally, organic hazes increase the atmospheric opacity at a broad wavelength range. This work will attempt to quantify the extent to which methane and haze complicate the detectability of surface biosignatures on Archean Earth-like analogues. The Archean Earth may have been covered in organic haze (Trainer *et al.* 2006, Kasting & Howard 2006, Arney *et al.* 2016), so it is important to consider haze when characterizing Archean Earth analogues. Organic hazes are formed from photolysis reactions involving methane, and may form on Earth-like planets when the atmospheric $CH_4$/$CO_2$ ratio is sufficiently high (Trainer *et al.* 2006, Arney *et al.* 2016, Arney *et al.* 2018). The *Archean* Proxima b runs from DG19 did not include haze formation in the atmospheres, though in several cases the $CH_4$/$CO_2$ and $CO_2$ mixing ratio were likely high enough to initiate haze formation (Meadows *et al.* 2018, also see Fig. 7 in Arney *et al.* (2018) for a similar haze formation parameter space). Hazes are generally highly opaque, so it is important to consider their effects on surface biosignature detection.

In this study, the haze which Meadows *et al.* (2018) simulated for Proxima b was added into the *Archean* DG19 runs "in post" in the observation simulator, because ROCKE-3D does not simulate haze formation. The atmospheric parameters of the DG19 *Archean* runs are not identical to those of Meadows *et al.* (2018), so self-consistent hazes in the *Archean* DG19 runs would be a little different from those of Meadows *et al.*, but should be a similar optical depth to the hazes that would form on the DG19 *Archean* runs to within an order of magnitude (see Fig. 7 in Arney *et al.* 2018). For quantifying the scattering and transmission features of the haze in the observation simulator radiative transfer calculations, we use the Khare *et al.* (1984) optical constants. The Khare *et al.* constants are used as a standard in the literature to model haze radiative effects because of their large wavelength coverage, despite their limitations. There has been some recent work in the exoplanet modeling community (e.g., Corrales *et al.* 2023) which has measured optical constant for hazes of more diverse composition, which may be more realistic for terrestrial planets. We hope that future work will explore the differences that updated optical constants would make on detectability.

Observing Proxima b would be a limiting case for the easiest exoplanet on which one could observe a biosignature due to distance considerations alone, because of its proximity to Earth at 1.3 pc. However, the majority of known exoplanets are much farther than this. Due to simple physics, a given object appears dimmer at farther distances. Thus, the question naturally arises of how detectable surface biosignatures would be at distances greater than 1.3 pc. In the present work, we treat this question only superficially, as a first step towards answering this question.

Observing small semimajor axis planets such as Proxima b becomes challenging for LUVOIR when the planetary systems are at great distances away from Earth. At great distances, very close-in planets have small star-planet angular separation, which makes them fall within the IWA of the coronagraph likely to be used for future space observatories (thus blocking out all their reflected light), at least at the wavelength range studied here. Proxima b is already at ~40 milli-arcseconds at maximum separation at 1.3 pc, so observing similar small systems at greater distances becomes challenging for coronagraphs. In the present work, we test the effects of greater distances on detectability just from the inverse square law. To do so, we place the DG19 Proxima Centauri run *Control* around a Sunlike star, with a semimajor axis of 1 AU (to achieve greater angular separation). We place the planet at 1 AU around a Sunlike star so that the planet is also in the habitable zone, but the incident starlight intensity at the TOA ends up being higher for this scenario, making the reflected intensity from this planet brighter than that orbiting an M-dwarf star (thus causing intrinsically higher SNR for the Sunlike case). Using this scenario, we test the detectability of a surface biosignature on *Control* at various distances from Earth. The Sunlike star at various distances does not correspond exactly to any real star in our galaxy at some distance from Earth. For this part of the work, we use the observation simulator as a lab to test the effects of distance on detectability (specifically, the SNR dropoff with the inverse square law of brightness). We do not explicitly tie our simulations to any mission's target star list. Future work should test detectability explicitly with the HWO target star list.

### *2.2 Spectral Observation Simulations*

The observation simulator used in this study is the Planetary Spectrum Generator (PSG), an online tool for simulating radiative transfer in planetary atmospheres (Villanueva *et al.* 2018). PSG can simulate observations over a broad wavelength range for a wide variety of atmospheric compositions and planetary geometries, and includes instrument noise models. See Villanueva *et al.* (2022) for details about PSG's capabilities and methodology. The 3D mode of PSG (Global Emission Simulator, "GlobES") ingests GCM output data, simulates the atmospheric radiative transfer at many latitudes/longitudes across the planet to generate reflectance spectra at the

TOA, and area-aggregates those spectra into a single, disk-averaged spectrum, just as an exoplanet would be observed with a distant observatory. The disk-averaged spectrum is hereafter referred to as the "planetary spectrum". For observation simulations with PSG, we used $a$ = 0.0485 AU (semimajor axis), $i$ = 90° (inclination, giving an edge on view of orbital plane). For our stellar spectrum model, and we chose parameters to fit the Proxima Centauri stellar spectrum from Meadows *et al.* (2018). An inclination of 90° was chosen to match the planetary mass prescribed by DG19, even though Proxima b has not been observed to transit and is not predicted to do so (Kipping *et al.* 2017, Gilbert *et al.* 2021). We simulate observations of reflection spectra while the planet is at quadrature (phases of 90° and 270°) to get the maximum star-planet separation, which is optimal for coronagraphic observations. Given that we choose an inclination of 90°, the eastern and western halves of the dayside (i.e., leading and trailing hemispheres of the planet) are fully in view at each phase observed.

The instrument parameters used in this work are from the LUVOIR A and B concepts (The LUVOIR Team 2019), taken from Checlair *et al.* (2021). In brief, LUVOIR-A and B are concepts for large (several meter aperture) telescopes with coronagraphs, optimized for observations in the visible-to-near-infrared. LUVOIR-A represents a best-case scenario of biosignature detection for the near future because of its larger collecting area, but is unlikely to be built in the near future due to its expected prohibitively high cost. The noise model used in the present work is a fairly simple model based on the LUVOIR-A/B instrument concepts, but provides a good order of magnitude estimate for how those instruments would operate. The knowledge gathered from the LUVOIR and HabEx studies will be used for development of HWO. The parameters for the instruments on HWO are still being defined at the time of this writing, but may be roughly similar to those of LUVOIR-B, as both are off-axis concepts. The LUVOIR-B concept has a mirror diameter of 8 m, whereas a main aperture of 6 m is currently being considered for HWO. The telescope aperture and other parameters of HWO are, however, highly uncertain at the time of writing and are being actively investigated. A smaller main diameter impacts the collection area of the telescope, but it also impacts the IWA of the coronagraph, and the ability of the coronagraph to block the star and preserve the planetary fluxes. The impact on the SNR by a smaller collecting area would be modest, but it can be substantial on the coronagraph throughput, and as such, the herein presented simulations can be thought of as a guide to what can be detected with the aperture and coronagraph designs as reported in the LUVOIR reports. As for the tentative instrument design of HWO as it is currently planned, Proxima b would lie within the IWA of the coronagraph with the system at 1.3 pc, making observations very challenging. However, changes to the instrument design may make Proxima b more easily observable by HWO, and there may exist future coronagraph technologies to alleviate the low throughput. We also tested simulated observations of Proxima b with HabEx (including a starshade), but found that such a configuration would not be suitable for observation of Proxima b, as the system's small star-planet angular separation makes it lie within the angular region blocked by the starshade.

We generate spectra in the VIS and NIR channels for both LUVOIR A and B to get the needed wavelength coverage to observe the spectral biosignature feature in question (which absorbs over 0.9–1.1 um). We considered the spectral resolving power as reported in the LUVOIR report for the VIS and NIR channels, which is 140 for the VIS channel and 70 for the NIR. The features explored in this paper are broader and wider than these resolutions, and since we compute the SNR integrated across all encompassing pixels, the SNR of the feature detection is, in principle, independent of the assumed spectral resolution. As such, the minimum spectral resolution needed to detect the biosignature absorption feature would simply be that required to spectrally resolve it from any nearby atmospheric feature (such as the water features ~50 nm away on either side). For integration time calculations, we assume that both channels can be used at the same time. In practice, a dichroic would be useful to observe this feature. However, depending on HWO's bandpass width, this feature may be possible to observe in a single observation.

*2.3 NIR Biosignature Feature*

The spectral biosignature under consideration in this work is a photosynthetic absorbance feature centered near 1016 nm (called the Qy feature in the photosynthesis literature), originating from bacteriochlorophyll *b* (Bchl *b*), the principal light harvesting pigment in an anoxygenic photosynthetic purple bacterium called *Blastochloris viridis* (formerly *Rhodopseudomonas viridis*) (Drews & Giesbrecht 1966). Figure 1 shows the *in vivo* reflectance spectrum in whole cells. The Qy absorbance of Bchl *b* in *B. viridis* is the longest wavelength absorption that is currently known for photosynthetic organisms on Earth.

Purple bacteria are thought to be among the first phototrophs to evolve on early Earth (Xiong et al., 2000), and geologic evidence suggests that they were present during the Archean Eon (e.g., Czaja et al., 2013). Today on the modern Earth, *B. viridis* is limited to shallow freshwater environments (Drews & Giesbrecht 1966). In the present work, we placed a *B. viridis*-like organism in the oceans of Proxima b, but this does not pose a major contradiction, because the photosystem and the organism's physiological tolerances are not strictly tied to each other. For example, chlorophyll *a* is used by cyanobacteria adapted to both saline and freshwater environments, as well as in land plants. Bchl b is also found in the halophilic, alkaliphilic, and thermophilic purple sulfur bacterium *Halorhodospira* (formerly *Ectothiorhodospira*) *halochloris* isolated from a soda lake (Imhoff and Trüper, 1977), which also has an absorbance peak at 1016 nm. We had *B. viridis* cultures readily available in the lab, so that is what was used in this study, but it otherwise has the same absorbance peak as its halophilic counterpart.

The photosynthetically active radiation (PAR) range for *B. viridis* is roughly 400 to 1035 nm (Kleinherenbrink *et al.* 1992). Photosynthetic life on Earth absorbs in the visible, the same spectral region as the Sun's peak energy output, and some extend the absorbance into the near-infrared. Photosynthetic activity is a function of locally available photosynthetic photon flux density (PPFD) *($\mu mol/m^2/s/\mu m$),* but it is also a function of energy per photon because only photons in a certain energy range are useable for photosynthesis (see review of basic processes in Schwieterman *et al.* 2018). The units of photon flux density ($\mu mol/m^2/s/\mu m$) are a measure of the number of photons available at a given wavelength. McCree (1972) proposed the definition of PAR units in terms of quanta (Einsteins or μmol photons $m^{-2}$ $s^{-1}$ $\mu m^{-1}$) rather than energy flux, because photosynthesis is a quantum process in which the relationship between photons absorbed and reactants (e.g. $O_2$ and $CO_2$) is stoichiometric. This definition gives the most consistent estimates of photosynthesis rate (and other biological processes) per unit of light flux across a range of light sources with very different emission spectra. The light harvesting pigment absorbance spectra are constrained by the available light spectrum (but this is not always the case), through both adaptation and acclimation, and therefore it has been suggested that alien photosynthetic organisms may adapt to best suit the output spectrum of their host star/local environment (Kiang *et al.* 2007, Schwieterman *et al.* 2018). The long wavelength NIR absorption feature of *B. viridis* serves as a model for photosynthesis adapted for life on planets orbiting M-dwarf stars, which emit more radiation in the NIR than in the visible.

The Qy absorption feature is close to the peak of a local maximum in photon flux density available at the surface of the DG19 Proxima b concepts, at least for *Control* and *Archean Low* (similar to *B. viridis*' local environment on Earth). This transmitting band of the atmosphere is bounded by two opaque water features. This NIR wavelength band would be an ideal absorption region for life on a water-bearing planet orbiting an M-star, because it is spectrally near the peak photon flux output of cool M-stars. Therefore, an analogue of the Bchl *b* pigment could be advantageous to use for photosynthesis on Proxima b, and it may be reasonable to expect that something similar would evolve in the photosystems of life on planets orbiting M-stars. Chlorin pigments are based on a molecular structure called a macrocycle — a large central ring created by joining together four separate five-atom rings (see Hoehler *et al.*, 2020 for an overview). The pattern of alternating double and single ("aromatic") bonds in the macrocycle, the ability of the nitrogen atoms to coordinate metal cations (e.g. Mg), and the conformational flexibility of the macrocycle and side chains allow for the unique biological function of pigments. Given the flexibility and variety of ways these macrocycles can be modified to tune their absorption, it's not unreasonable to expect that a pigment-type structure could evolve on another planet.

To detect this surface biosignature in practice, an astronomer could observe the star and the planetary reflection spectra. They could use these to generate a model of the incident starlight spectrum at the surface of the planet. Astronomers could then look for photosynthetic surface biosignatures near the peaks of the incident photon flux density available at the surface. If an unidentified absorption feature was observed near such a peak, it may be reasonable to suspect the presence of photosynthetic life on this planet, but other inorganic absorbing species on the surface and in the atmosphere would have to be ruled out. Molecular constraints may limit the longest wavelength useful for photosynthesis, such that the pigment peak absorbance may not exactly match the peak in irradiance. For example, the far-red chlorophyll *d* absorbance peak occurs a little shortward of the peak photon flux wavelength in its spectral environment on Earth (Larkum & Kühl, 2005), due to molecular constraints that would cause tradeoffs in quantum yield if extending charge separation to longer wavelengths (Viola *et al.* 2022). In the present work, the *B. viridis* Qy band being shortward of the Proxima Centauri peak emission could be an example of a similar case.

We measured the reflectance spectrum of a pure culture of *B. viridis* DSM 133 (type strain) in growth media with a field spectroradiometer (Analytical Spectral Devices). The spectral range of the measurements was 0.35–2.5 µm. The culture was in a Petri dish on top of a black background and illuminated with a 200 Watt incandescent bulb, and the contact probe was positioned 5 cm above the culture. The reflectance spectrum is a function of the density of the bacterial cells. The culture measured in Figure 1 was grown in conditions achieving densities higher than typical in natural habitats on modern Earth, but not as dense as possible with greater resources or enhanced growth conditions.

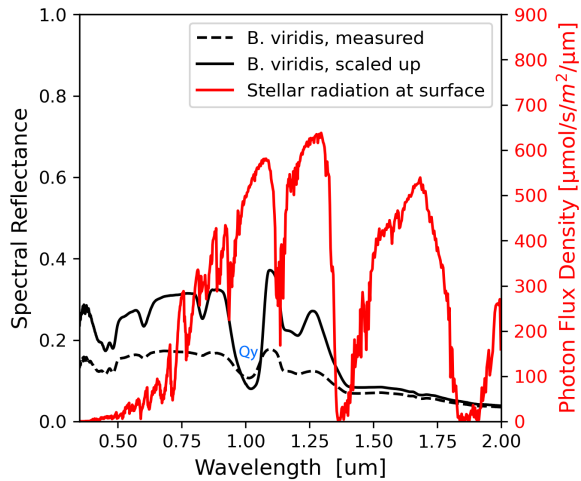

**Figure 1:** (dashed black) measured reflectance spectrum of *B. viridis*, and (solid black) scaled up version of this spectrum with the Qy biosignature feature having similar contrast to the vegetation red edge. The latter is the base-case spectrum used to represent a high density surface biosignature in this work, which is then linearly diluted in concentration in detectability simulations. Overplotted (red) is the dayside average incident starlight spectrum available the surface of the *Control* Proxima b concept, as modeled with PSG. The Qy feature overlaps with a large peak in incident radiation at the surface, which would make this pigment well adapted for photosynthesis on such an M-star hosted planet. The data behind this figure is available in machine readable format.

In GlobES, the biosignature reflectance spectrum and the grid cell albedo are mixed together using linear areal mixing (see PSG handbook section 6.7 for details) using the biological organism's abundance (area cover fraction) at that grid cell. Using this approach, we needed the biosignature's reflectance spectrum to be equivalent to that at maximal cell density, which we then linearly dilute in concentration to test the detectability of this biosignature at various abundances. This represents a realistic dilution of the pure culture (~$10^9$ cells/mL) to cell densities found in the natural environment ($10^5 - 10^6$ cells/mL in a variety of lacustrine and marine settings). We start with the maximally scaled up spectrum to establish a limiting case for best chances of any type of surface biosignature detectability. Scaling down the area concentration in PSG may alternatively be viewed as linearly scaling down the biosignature absorbance spectrum itself (so the spectral feature is weaker, more akin to the absorption of *B. viridis* itself).

To accomplish our spectral initial scaling, we devised a general spectral scaling scheme to estimate the biosignature's reflectance for different potential cell densities. The reflectance spectrum of a cell culture or population will not scale with cell density as a simple scalar multiple; instead, absorption by pigments would decrease reflectance in relevant bands, which would contrast with increasing reflectance by scattering by membranes and turbidity in the medium in other bands. This is demonstrated for reflectance spectra of cyanobacteria at different dilutions by Battistuzzi *et al.* (2020). In lieu of having cell cultures at different densities and grown under different conditions, or a proper radiative transfer model to account for changes in content of absorbing and scattering elements, we scaled up the measured reflectance spectrum to emulate how a cell culture would increase its absorbance by various features associated with Bchl *b* (at maximum cell density) and carotenoids (anti-oxidant pigments that are always present in photosynthetic organisms on Earth) and how the culture would increase reflectance at other wavelengths. This rescaling was constructed based solely on expert judgment to produce hypothetical spectra with a stronger signal at the Qy band of Bchl *b*. We limited the scaling to one that achieves a contrast of ~0.2 at the Qy band (spectral contrast = continuum – trough), which is comparable to values for the vegetation red edge, though we have no basis to expect *B. viridis* itself to be able to achieve such high spectral reflectance. We intensified the Qy feature so that we could use this "maximum cell density" case to calculate a lower bound for integration time to detect this surface biosignature on Proxima b. Being that this scaled up reflectance spectrum differs from actual absorbance of *B. viridis*, we henceforth refer to the hypothetical organism in this work as a *B. viridis*-like organism ("BVLO"). We attempt to detect this hypothetical organism's light absorption on Proxima b while the organism is present at various area concentrations.

### *2.4* **B. viridis**-*like Organism Surface Abundance Distributions*

The BVLO reflectance spectrum was distributed over the surface of the planet using a simple function that results in a 2D array which yields the BVLO surface area coverage fraction of any given grid cell of the GCM run surface. We choose to parameterize the BVLO abundance as an area coverage fraction for simplicity of calculation in PSG, rather than having to alter the spectrum across the planet. In effect, the area fraction represents a BVLO spectrum dilution factor. This is because at the subgrid scale, surface spectral components are mixed together linearly in PSG, weighted by their areal coverage fraction, producing a single surface reflectance spectrum for each grid cell, as would be observed in reality for areas below the observation's spatial resolution limit. The simple growth response function for parameterizing BVLO abundance (how well the organism grows as a function of various parameters) that we adopt is:

$$A_{bac} = A_{max} \cdot f_{ow} \cdot f(T) \cdot f(PAR) \tag{1}$$

Where all terms range from 0-1. This simple method of parameterizing organismal abundance with multiplicative functions is generalizable to any organism by changing the growth response function, and can be used on any GCM run.  is the areal cover fraction of the BVLO at a given grid cell (called as such to be reminiscent of bacterial abundance, though the BVLO may be thought of as a more generic organism functional type). In general, areal coverage fractions do not precisely represent biological organism distributions, so it is better to think of  as a spectral dilution factor in each grid cell. Additionally, there are many reasons why the light absorption of a surface component would increase (other than increasing area cover fraction, e.g. by increasing biomass concentration). Therefore, the cover fraction  will henceforth be referred to as an "abundance," which is a more general term.

$A_{max}$ is a scaler that takes on a constant value across the whole planet, which we vary across a range to test the integration times required to detect various global BVLO abundances. In other words, all parameters of equation 1 are intrinsic to the climate state besides , which we vary to scale the global abundance up and down (illustrated in figure 3). $f_{ow}$ is the open water fraction at any grid cell, which is a variable directly output by ROCKE-3D. The BVLO are placed throughout the open ocean to test detectability of planet-wide phytoplankton-like organisms. The nightside of the planet also has open water in these ROCKE-3D models, but there is no light for photosynthesis, so we take the BVLO abundance to be negligible on the nightside. $f(T)$ is a temperature growth response function, and $f(PAR)$ is a PAR growth response function. The $f(T)$ and $f(PAR)$ used in this work are plotted in

Figure 2. In this work, we take the BVLO's abundance (as correlated to growth rate) to be dependent on the annual mean sea surface temperature, which is an output of ROCKE-3D. $f(T)$ is taken from Yan & Hunt (1999), and is given by:

$$f(T) = \left(\frac{T_{max} - T}{T_{max} - T_{opt}}\right)\left(\frac{T-T_{min}}{T_{opt} - T_{min}}\right)^{\left(\frac{T_{opt} - T_{min}}{T_{max} - T_{opt}}\right)} \text{ for } T_{min} \leq T \leq T_{max}, 0 \text{ elsewhere} \quad (2)$$

Where T is the sea surface temperature at a given grid cell, $T_{min} = -10°C$, $T_{opt} = 10°C$, and $T_{max} = 20°C$. We chose these values to parameterize the BVLO's temperature-growth response to mirror that of terrestrial cryophiles/psychrophiles (Russel *et al.* 1990), so the BVLO would be well adapted to life in the ocean on the DG19 Proxima b concepts (another significant difference from *B. viridis* itself). Here and hereafter the word "terrestrial" is taken to mean "of the Earth," rather than "on land." The sea surface temperatures on the DG19 runs under study in this work range from -2.1 °C ≤ T ≤ 8.8 °C, so the BVLO is assumed to be well adapted to its environment. It seems reasonable to expect that alien organisms would be as well adapted to their native environment as possible, though there may exist some intrinsic physical/kinetic limitations due to enzymatic functions.

Our parameterization for $f(PAR)$ is given by

$$f(PAR) = \tanh\left(\frac{PAR \cdot ln3}{2k}\right) \quad (3)$$

Where PAR is the PAR flux incident at the surface at a given grid cell and $k$ is the value of PAR at which $f(PAR) = 0.5$, chosen in this work to be 150 $\mu mol/m^2/s$. This parameterization provides an approximately linear response to light intensity over 0 to ~300 µmol/m²/s, saturating at about 600 µmol/m²/s. The saturating value implies this organism can reach maximum density of cover at light levels that are high for typical purple bacteria on Earth today, but comparable to the mid-latitudes visible radiation on Earth and to what is optimal for some cyanobacteria (see Ritchie, 2008, for a comparison of light saturating responses of different phototrophic bacteria). Since our organism dominates at the surface of the planet, it can be expected for it to be adapted to the surface light. The maximum PAR on any of the DG19 runs under study in this work is $683.6 \, \mu mol/m^2/s$ (fairly low compared to terrestrial high-light environments), so the effects of photoinhibition are not included in the parameterization of $f(PAR)$.

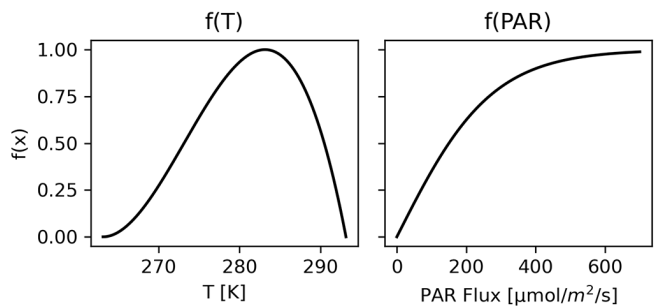

**Figure 2:** $f(T)$ and $f(PAR)$ plotted over the range of values that occur in this work. These functions are used calculate the parameters in equation 1 at every grid cell to yield $A_{bac}$.

Global area-weighted average BVLO abundance is given by:

$$\overline{A_{bac}} = \frac{\sum_{i,j}(A_{bac\,i,j} \cdot Area_{i,j})}{\sum_{i,j} Area_{i,j}} = \frac{Total\ Area\ of\ BVLO}{Total\ Area\ of\ Planet} \quad (4)$$

Where $i$ and $j$ are latitude and longitude grid indices. $\overline{A_{bac}}$ is a useful parameter to analyze results with, as it represents total global BVLO area coverage. Thought of another way, this number is a proxy for how much the BVLO spectrum has been diluted over the whole planet.

For the *Archean* scenarios with haze added, the BVLO abundance maps are identical to those used for the *Archean* scenarios without haze. In reality, haze would block out some PAR and cool the sea surface temperature (thus changing the BVLO abundance distribution by equation 1), but without rerunning the GCM it is difficult to quantify these climate effects. Therefore, hazes are included in this work purely to quantify the changes in surface biosignature detectability due to changes in atmospheric opacity.

## 2.5 Detectability Calculations

To calculate SNR for a planetary spectrum at a given $A_{max}/\overline{A_{bac}}$, we first used GlobES to generate spectra of each DG19 GCM run with BVLO present on the surface. Then, to test the detectability of the Qy feature, we put the BVLO on the planet in the same abundance but instead of using the BVLO spectrum, we used the same spectrum but without the Qy absorbance feature. To construct this BVLO spectrum without Qy, we replaced the Qy feature with a straight line connecting the points at 888nm and 1093 nm of the measured spectrum. With this method, it is possible to calculate the difference in planetary spectral albedo due to the Qy feature alone. Then, one can calculate the integration time needed to detect the Qy feature in the planetary spectrum (i.e., the integration time such that the noise would be low enough to distinguish the Qy from the local continuum).

In general, adding the BVLO to the DG19 runs increases the planetary albedo significantly (compared to the case with no BVLO at all), as the BVLO has high reflectance and is replacing water, which has low albedo. This increase in albedo alone due to the BVLO is not interpretable as a biosignature (as something as simple as larger ice surface coverage would cause the same thing, i.e., the signals are degenerate), so integration times to detect this broadband albedo increase are not calculated (especially considering the observational degeneracy between exoplanetary radius and albedo). Rather, we focus only on the detectability of the Qy absorption feature. If a surface map retrieval such as proposed by Berdyugina and Kuhn (2019) revealed a region of anomalously high spectral albedo where there was open water expected, this might be cause for further investigation, but it would not be interpretable as a biosignature by itself.

We use PSG to simulate various noise sources, including noise from reads and dark frames, thermally induced noise, etc. We note in particular that the artifacts due to exozodiacal dust are modeled as a noise source, where we assume a scale factor of 4.5 relative to the amount of dust in our Solar System. We do not explicitly include the noise effects of stellar speckles, but assume a coronagraphic performance of $10^{-10}$, following the LUVOIR Final Report (2019). See Villanueva *et al.* 2022, PSG handbook Chapter 8, for details on noise modeling. The biosignature under investigation spans a spectral range in which the noise transitions from being source dominated to detector dominated . The noise in the NIR channel is much greater than the VIS channel because of high read and dark noise from the NIR detector. This difference is due to the different material properties of the detectors (charged coupled detectors in the UV-visible versus HgCdTe in the near-infrared). We model our observations such that a CCD is used shortwards of 1 um, and a HgCdTe detector is used longwards of that.

The SNR is a critical parameter for observations of any target which describes how good the observational data is. In this work, it is calculated as follows (adapted from Checlair et al. 2021):

$$SNR_\lambda = \frac{I_\lambda(\text{w/ Qy feature}) - I_\lambda(\text{w/o Qy feature})}{N_\lambda} \qquad (5)$$

Where $SNR_\lambda$ is the SNR at a given wavelength, $I_\lambda$ is the value of the spectrum at that wavelength (in any unit), and $N_\lambda$ is the noise at that wavelength.

First, SNRs are calculated for an arbitrary integration time (in this case we used an integration time of 1 hour for the initial spectra generation). Using the SNR calculated for that arbitrary integration time, one can calculate integration times required to achieve the desired SNR. In this work, we use an SNR of 5 as a minimum threshold for detectability. The integration time required to achieve an SNR of 5 is calculated as follows:

$$t = t_0 \frac{5^2}{\sum_\lambda (SNR_\lambda)^2} \qquad (6)$$

Where t is the integration time and $t_0$ is the initial exposure time of 1 hour. In this work, an integration time of 1000 hours is considered to be the maximum integration time for detectability, following Checlair et al. 2021. This is assumed to be the maximum amount of telescope time that would ever be allocated to observing a single target of major scientific importance. In reality, any given single target would probably be observed for less than this, especially for first round of observations. The 1000 hour mark could possibly be realistic for follow up observations of extremely high value targets. Hereafter, the integration time required to achieve an SNR of 5 is referred to as "the" integration time.

### 3. Results

#### 3.1 BVLO Abundance Maps

The various subcomponents of equation 1 each generate unique arrays and multiply together to generate BVLO abundance maps, which is illustrated in Figure 3.

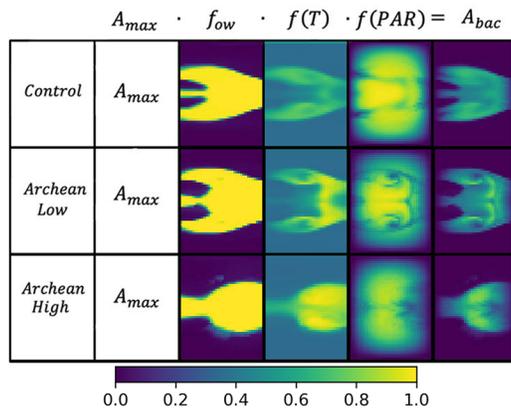

**Figure 3:** Array components of equation 1, the function which generates BVLO abundance maps, shown for the dayside of the planet only. The 2D arrays are multiplied together to yield the BVLO abundance array, shown in righthand column. Colorbar scale is for an $A_{max}$ of 1 (the global dilution scaler).

The BVLO abundance maps are plotted in Figure 4, along with some additional diagnostics which illustrate the global distributions. The eastern halves of the daysides of all DG19 states contain more BVLO than the western halves when biological activity is parameterized in this way. This hemispherical asymmetry is because on these GCM runs, warm waters tend to be located near the equator on the eastern halves of the daysides.

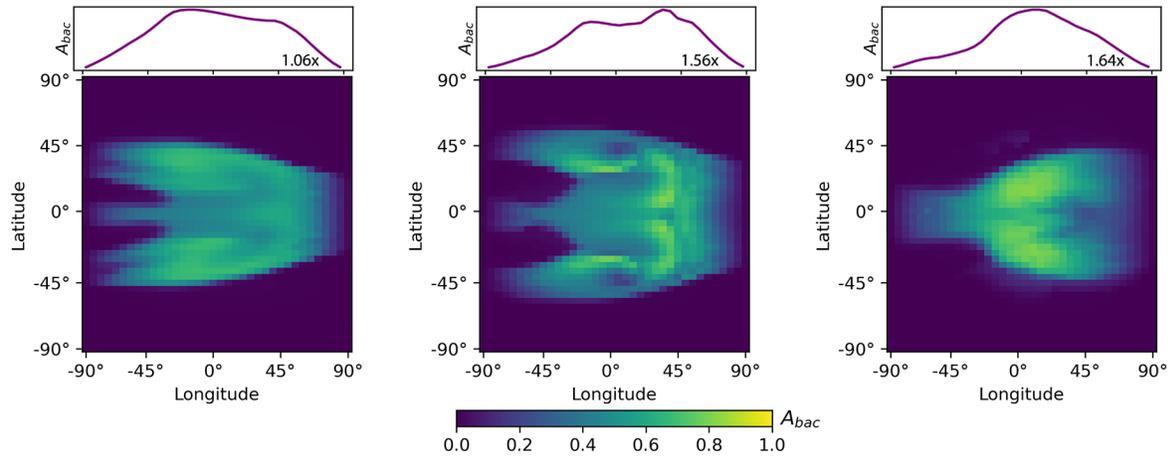

**Figure 4**: *B. viridis*-like organism (BVLO) abundance maps on the dayside of DG19 Proxima b states ($A_{bac}$ as a function of latitude and longitude), shown for $A_{max} = 1$. Plotted above the maps is the surface area coverage fraction in each longitude band. The numbers at bottom right of these top panel plots are multipliers indicating the total hemispherical east:west BVLO abundance ratio. When observing this planet at a phase of 270° (for an inclination of 90°), only the eastern half of the dayside is visible, and only the western for a phase of 90°.

### *3.2 Simulated Spectra*

A few of the spectra used to calculate integration times are shown in Figure 5. The spectra show the Qy biosignature feature, as well as distinct methane and water vapor absorption features.

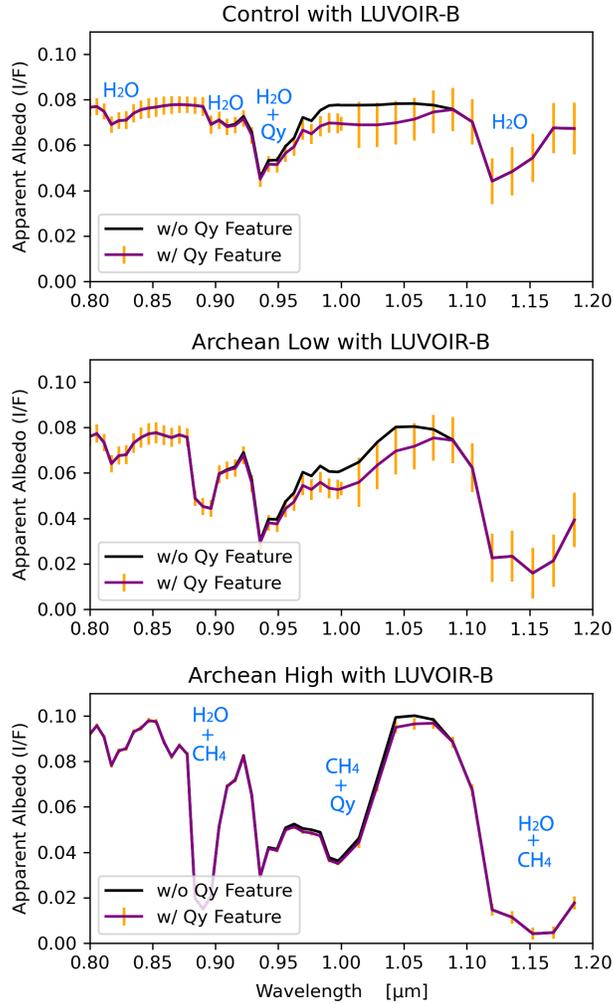

**Figure 5:** Simulated planetary reflectance spectra from PSG for all DG19 states at a phase of 270°, observed with LUVOIR–B. Plotted are (purple) the planetary spectra with the BVLO, and (black) the planetary spectrum with the modified BVLO spectrum without the Qy feature. Noise is shown by orange error bars. One can see that the Qy absorbance feature decreases the spectral albedo in the $0.9 - 1.1$ μm range. Spectra with $A_{max} = 1$ are shown, so as to best illustrate the spectral effect of the Qy feature. The integration time used for these spectra is that which would bring the overall SNR to 5 (values given in figure 7). Note the increase in noise longwards of 1um is due to a transition to a HgCdTe detector.

Figure 5 shows the methane absorption features getting progressively stronger as methane concentration increases from *Archean Low* to *Archean High*. The water vapor features remain roughly constant across all three scenarios. Control and *Archean Low* have a similar amount of signal (difference in the spectra without Qy and with Qy). The most useful signal comes from the spectral range ~$0.97 - 1$ $\mu m$ (this is where the noise is smaller than the difference of planetary spectrum with/without Qy). The water feature near $0.95$ $\mu m$ blocks out some of the signal. Longwards of 1 $\mu m$, the simulated instrument detector transitions to a different mode of observation for the NIR band, which has comparatively high noise due to the effects of detector material properties.

The bottom panel of Figure 5 shows that the signal is very small for *Archean High*, thus requiring extremely long integration times to obtain an SNR of 5. This small signal is because of the large amount of methane in the atmosphere making the atmosphere very optically thick to light in the Qy spectral region, so surface absorption signatures are less evident in the overall planetary spectrum.

In Figure 6, we show the spectral confusion that is possible with the Qy feature due to methane. Methane strongly absorbs around 1 µm, but this degeneracy with the Qy feature can be broken by observing at other wavelengths where methane is known to absorb. If the Qy feature is present, then there would be extra absorption at 1 µm that would not fit the methane absorption at other wavelengths. In other words, one would have to observe at least one other methane absorption band and perform a fit using a spectral library and some kind of Monte-Carlo retrieval approach, and then one should be able to rule out methane as causing all the absorption at 1 µm. It may be difficult to attribute this extra unexplained absorption to life, but perhaps there would be additional biosignatures observable from this planet (atmospheric or surface) to support the interpretability. In general, to interpret this Qy feature as a biosignature, one would have to rule out all other known abiotic atmospheric and surface spectral signatures. This problem could be approached in a similar fashion as shown in Figure 6, by observing at other wavelengths, but ultimately the interpretability may remain ambiguous because we have no way of knowing what spectra of alien life will look like. In other words, alien life may use pigment molecules that absorb at unknown (and hard to predict) wavelengths.

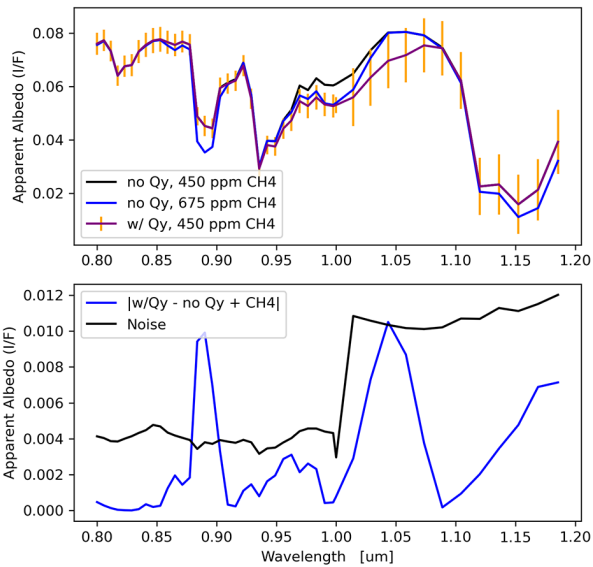

**Figure 6:** Spectral confusion of methane and the Qy feature, shown for the *Archean Low* scenario. Top: (purple) the planetary spectrum for *Archean Low* is shown with the Qy feature (with noise, integration time of 1 hour). Also shown are: (black) *Archean Low* without the Qy feature, and (blue) Archean Low without the Qy feature, but with extra methane. Adding extra methane mimics the Qy absorption feature around 1 µm.
Bottom: (blue) signal/spectral difference between "*Archean Low* with Qy" and "*Archean Low* without Qy, plus extra methane," overplotted with (black) the noise from "Archean Low with Qy". For most of the spectral range, the difference would be undetectable at this integration time (meaning, that the signal < noise). However, the signal is greater than the noise around 0.87 µm, so the degeneracy could be broken by observing at this wavelength.

### 3.3 Integration Times

Figure 7 shows the integration times necessary to achieve an SNR of 5, for all three DG19 scenarios under consideration in this work. Integration time increases super-exponentially with decreasing abundance. Table 2 shows the approximate detectability thresholds (minimum abundance required for detection) for each DG19 scenario. For moderate to high abundances, the biosignature is detectable on *Control* and *Archean Low* in a matter of some tens of hours. For *Archean High*, hundreds of hours is a more typical integration time.

To facilitate the interpretation of the results of Figure 7, one can divide the $\overline{A_{bac}}$ by the scenario's $f_{hab}$ from Table 1 to derive a globally averaged linear dilution factor for the BVLO reflectance spectrum, where it is present. This linear dilution factor is a multiplier which represents the amount by which the BVLO spectrum was diluted from the maximum density base case. This dilution factor allows quick comparison between a given diluted BVLO spectrum and the contrasts of spectral features that occur in terrestrial biology. For example, the BVLO becomes undetectable on *Control* at an $\overline{A_{bac}} \approx 0.01$, which corresponds to a dilution factor of 0.024. The BVLO reflectance spectrum can be multiplied by this factor, and then the contrast of the Qy feature from that spectrum can be compared to contrasts observed in remotely sensed biological spectral features. This comparison will give an idea of what biological densities and functional types would be detectable if the BVLO were similar to terrestrial biology. We note that such comparisons of plausible alien life to terrestrial biology should be carried out with caution, as there are many uncertainties with extraterrestrial biology and its pigments (supposing they exist), such as cell density and intrinsic pigment absorption strength. The spectral contrast of an absorbance feature, when an alien organism is spread over large areas, could be quite different from what we have on Earth. With caution in mind, there exists a rough analogue on the modern Earth to the widespread oceanic BVLO scenario which we setup on Proxima b. NIR-absorbing pigments of aerobic anoxygenic phototrophs are found globally distributed in coastal areas and in the open Indian, Atlantic, and Pacific oceans and can comprise significant fractions of the total bacterial cell numbers (Kolber et al., 2000; Jiao et al., 2007). However, the remote detectability of this type of organism (or its NIR-absorption feature's spectral contrast over large areas) is yet to be determined. It would be a useful area of future work to compare this NIR feature's remote detectability, and that of other organisms, to the spectral contrasts in present work.

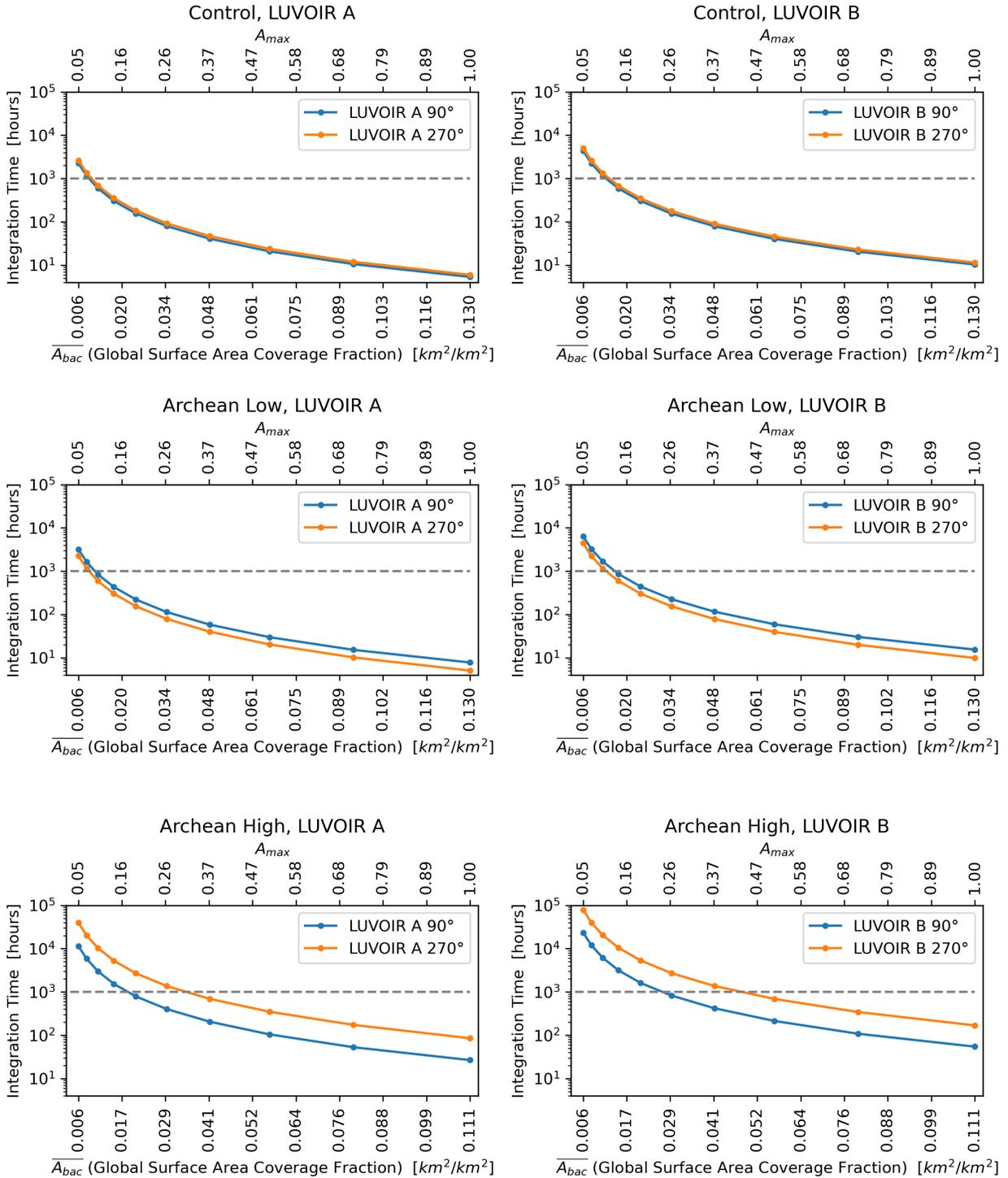

**Figure 7:** Integration times required to achieve an SNR of 5 with LUVOIR-A and B, plotted as a function of $\overline{A_{bac}}$ and $A_{max}$. $A_{max}$ shows the peak value for BVLO area coverage fraction/spectral dilution factor possible in any grid cell, whereas $\overline{A_{bac}}$ shows the global average coverage fraction/spectral dilution factor. The gray dashed line shows the detectability threshold criteria of 1000 hours. At intermediate abundances for the less opaque atmospheric

scenarios (*Control* and *Archean Low*), the BVLO Qy feature is detectable in some tens of hours.

**Table 2**
Global Cover Fraction ($\overline{A_{bac}}$) Detectability Thresholds

| DG19 Scenario Name | LUVOIR-A | LUVOIR-B |
|---|---|---|
| Control | 1% | 1.4% |
| Archean Low | 1.1% | 1.55% |
| Archean High | 2.6% | 3.7% |

Note: These values represent the minimum $\overline{A_{bac}}$ to be detectable (i.e., the $\overline{A_{bac}}$ required to achieve an SNR of 5 in 1000 hours of integration time). These numbers are phase averaged, and shown as a percent of total planetary surface area cover/spectral dilution factor.

The presence of methane decreases the detectability of this biosignature because of the increased atmospheric opacity in the $0.95 - 1.05$ $\mu m$ range. This increased atmospheric opacity is part of the reason why the scenarios with more methane require a longer integration time to achieve a given SNR (for *Archean High*, this is the dominant reason). Clouds and surface albedo also play a large role in detectability, especially in phase asymmetries of detectability.

Although all scenarios contained more BVLO on the eastern halves of the daysides (observed at a phase of 270°), Figure 7 shows that for *Control* and *Archean High*, the biosignature is easier to detect at a phase of 90°. This is because of a higher albedo at a phase of 270° (in either case due to surface components or clouds), which leads to more total noise. The eastern half of *Archean High*'s dayside is much cloudier than the western half, so the phase differences in detectability are dramatic (mostly due to decreases in atmospheric transmission).

Organic haze increases the integration time necessary to detect this biosignature, but only by a small factor. For all scenarios tested (*Archean Low* and *Archean High* at both phases, with LUVOIR A/B), the $\frac{\text{Integration time with haze}}{\text{Integration time without haze}}$ ratio was about 1.4, with the full range being from $1.3 - 1.5$. Therefore, the haze opacities tested in this work generally do not preclude detection of this biosignature. Other hazes may be extremely opaque, but the Meadows *et al.* (2018) haze used in this work has only a moderate effect on detectability, being relatively optically thin compared to organic hazes formed under other $CO_2/CH_4$ mixing ratio conditions.

Figure 8 shows the integration time required to detect the Qy feature on *Control* ($a$ = 1 AU) when the system is located at various distances from Earth. *Control* is the clearest atmosphere among all the DG19 scenarios tested, so this test represents a best-case scenario for detectability of surface biosignatures on a Proxima b-like planet. Because of the brighter Sunlike star, the detectability at 5 pc is comparable to the detectability at 1.3 pc using an M dwarf star. However, integration times quickly become prohibitively long at further distances. By 15 pc, the Qy feature is undetectable with LUVOIR-B. Judging by these results, surface biosignatures may be detectable on exoplanets within 10-15 pc, depending on the organismal abundance/spectral contrast. While the results of this simple test are somewhat illuminating, future work is needed to rigorously examine the detectability of surface biosignatures on more distant planets.

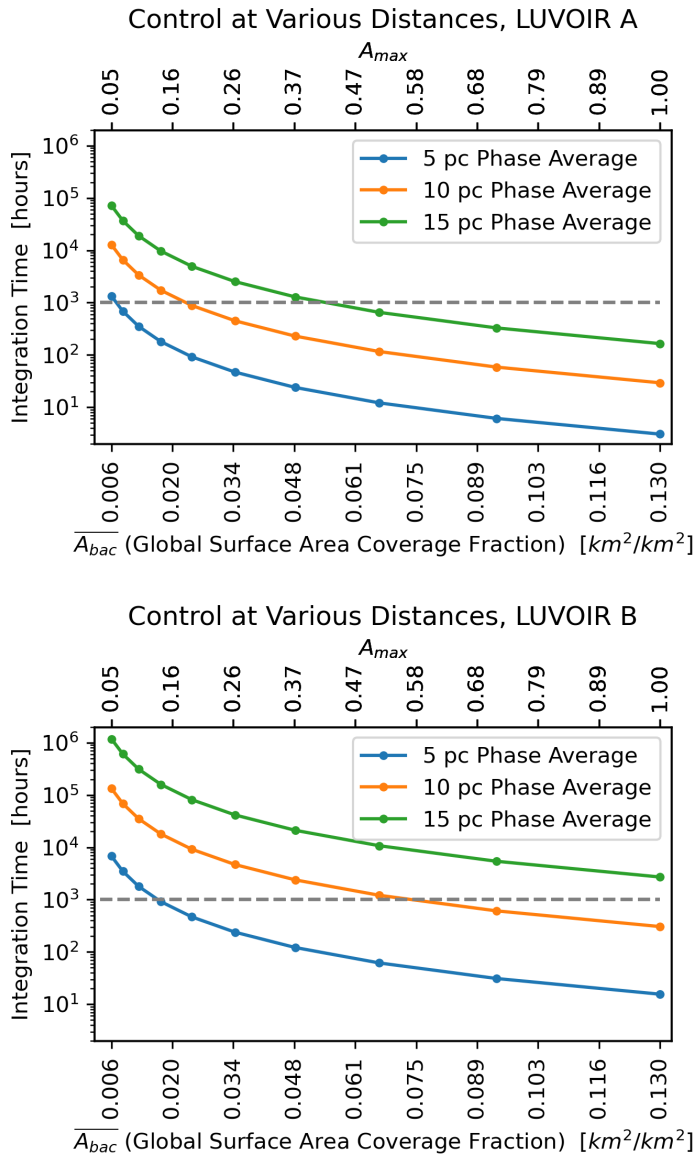

**Figure 8:** Integration times for *Control* at various distances, placed in orbit around a Sunlike star at 1 AU. The biosignature is completely undetectable at 15 pc when using LUVOIR-B, but feasibly detectable otherwise. $a = 1$ AU was chosen because at distances farther than 1.3 pc, larger angular separation was needed than Proxima b has in reality (so that the planet is not blocked by the coronagraph), and a Sunlike star was chosen to be roughly climatologically consistent with the DG19 states.

Figure 9 is included as a visual summary of the present work. The images in this figure were made by using GlobES to generate spectra at the TOA of every gridpoint of the GCM state, which are then converted into an RGB image. Images such as these can be created for any GCM state. Details of image generation are covered in Kofman *et al.* 2024 (in review).

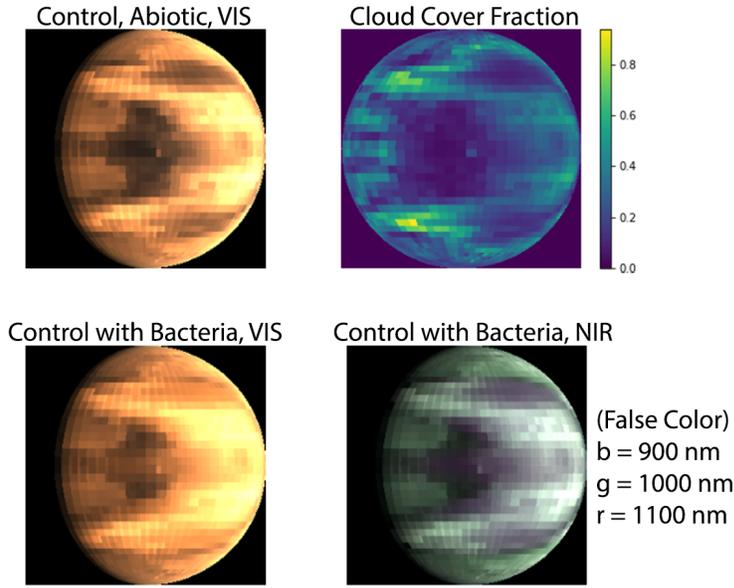

**Figure 9:** "Hyper-realistic" RGB images of *Control* at a phase of 40°, generated with PSG GlobES. Clockwise starting from the top left: (a) true color image, without BVLO. Proxima b appears orange because Proxima Centauri is a cool star which emits mostly red light in the visible. In this image, the darkest regions on the planet are oceans, and the lighter regions are ice or clouds. To identify clouds, compare with b. (b) the cloud cover fraction, with high numbers indicating grid cells mostly covered with clouds. (c) false color image of *Control* with BVLO. The BVLO appear purple because of Qy feature strongly absorbs in the green channel at 1 $\mu m$. (d) same as a, but with BVLO added ($A_{max} = 1$). The BVLO fill in the open water regions. To easily see regions of BVLO coverage, compare with purple regions in c.

## 4. Discussion

The model we used to parameterize BVLO abundance (equation 1) is very simple and idealized. Most notably, our model lacks a nutrient dependency because nutrient parameterizations were not included in the DG19 runs. A proxy for nutrient concentration would be sea-surface height (because upwelling waters are often nutrient rich), which could be included in future work without the need for nutrient parameterizations. On Earth, most phytoplankton live near the continental coasts where there are high nutrient concentrations due to runoff and upwelling.

Terrestrial phytoplankton concentrations/abundances do not reach such high densities as the highest abundances we tested in this work. Thinking of $A_{bac}$ as a spectral dilution factor in each grid cell, the BVLO Qy spectral contrast at the higher end of the $A_{max}$ range tested in this work is too large to accurately represent oceanic microbe reflectance, but such a large spectral strength is seen on Earth from land plants. What's more, the $\overline{A_{bac}}$ at these high $A_{max}$ are realistic for land plants on Earth. Thus, the present work suggests that organisms with land plant-like high area coverage and spectral contrasts may be detectable on nearby exoplanets. Future work is needed to rigorously test this result using GCM runs from land planets with various continental configurations. Roughly speaking, the intermediate abundances/spectral strengths we tested in this work may be analogous to very densely packed terrestrial microbes (whether aquatic or on land/coastal, such as bacterial mats). The lower $A_{max}$ values tested in this work may be a reasonable approximation of spectral contrast of phytoplankton-like organisms. At these low $A_{max}$, the BVLO spectral signature was undetectable on Proxima b. Therefore, direct spectral detection of diffuse microbial pigments may be difficult or impossible for all exoplanets, but future work is needed to test various types of spectral features to confirm or deny the generality this hypothesis. Future work is also needed to rigorously test the tentative conclusions drawn from Figure 8 in this work. Pigments with a diversity of spectral features should be tested on various types of planets and at varied distances from Earth.

Simulating feedbacks between the organism distribution and the atmosphere are outside the scope of this study, and this is itself a research challenge in Earth Science and global climate change (Fisher & Koven, 2020). Uncoupled modeling approaches usually are conducted to simulate biota distributions with offline climate forcings, or prescribe the biota changes to evaluate impacts on the atmosphere. A model of metabolic fluxes of an Archean $H_2$-based marine ecosystem of 3 primary producers (methanogens, anoxygenic phototrophs, acetogens) was implemented in a 1D model with a stagnant ocean surface boundary layer coupled to a radiative-convective/photochemical model for the atmosphere but without interactive climate (Kharecha et al. 2005); these coupled models were used to constrain Archean atmospheric CH4 profiles and substrate fluxes necessary for the different metabolisms. In our study, we took the approach to estimate anoxygenic phototroph distributions given the climatological constraints on biophysical controls (temperature, water, light). Chemical nutrient models would be a particularly relevant area of ecosystem limitations to consider in future work. Conducting coupled ecosystem-ocean-atmosphere simulations in order to capture the biophysical feedbacks and the biogeochemical in determining biogeography, climate, and chemistry, is a rich area for future major work involving significant model development.

Although the eastern halves of the daysides of these aquaplanets from DG19 were more populated with BVLO via equation 1, the BVLO was not necessarily easier to detect when observing that planetary phase alone. This is primarily because of cloud distributions. Considering the effects of clouds, there seems to be no preferred aquaplanet hemisphere for observing surface biosignatures.

A more realistic radiative transfer model for the bacterial reflectance would improve the present work. Parameterizing the BVLO abundance with a linear dilution factor in each grid cell does not allow for easy interpretation of biological abundances or cell densities. Aquatic microbes reflect light from the ocean column at some depth, not just the surface as parameterized in the present work. Therefore, interpreting $A_{bac}$ as a coverage fraction is not fully correct (at least not over water areas. This parameterization would work slightly better over land). It would be important in future work to more carefully parameterize reflectance as a function of microbial cell density than is done in this work, so that results are more easily interpretable (besides from just a spectral contrast/dilution perspective).

The instrument parameters for HWO are likely to change over the coming years as the development proceeds. Using the parameters for LUVOIR-B is only meant as a rough approximation for detectability with an HWO-class telescope. Importantly, HWO's current design considers a 6 m main aperture, which relative to the 8 m considered for LUVOIR-B, would have a substantial negative impact on the coronagraph throughput at 1 um for Proxima b as proposed here, hindering its detection. As the parameters for HWO mature and advance over the design phase, the detectability of these features would have to be recalculated for the evolving telescope and coronagraph considerations.

We note that it would have been most ideal to match the solar constants (incident flux at the TOA) of the 1 AU case to that of Proxima Centauri b, so that the detectability experiments could be more easily compared. However, the 1 AU case was only meant as a quick test, so this wasn't accounted for.

If surface biosignatures are a desired science goal of HWO or other future telescopes, the results of this work imply some suggestions:
1. Considering the hypothesis of spectral overlap of photosynthetic absorption and star photon flux emission, perhaps HWO's detector should be optimized for the VIS, as habitable-zone planets around mostly NIR emitting stars are likely to be too close-in to their stars to observe at typical exoplanet distances (because photons from the planet would be blocked by the coronagraph). In other words, the star's peak photon flux emission wavelength moves shortwards as the star's habitable zone moves outwards, and we should look for surface biosignatures at wavelengths adapted to the host star/planet.
2. A large bandpass is particularly desirable for surface biosignatures, as many of these features tend to be spectrally broad (particularly the Qy feature used in this work, at almost 200 nm width). A dichroic or similar apparatus may be beneficial. The need to observe at multiple wavelengths to rule out degeneracies furthers the benefit of a large bandpass.
3. Long integration times will likely be necessary, especially for more distant planets than Proxima b. A strategy for optimizing repeated observations would be needed. Perhaps a surface biosignature could become a feasible observation goal only after a target of significant interest is identified. For example, if a habitable planet with abundant water vapor and $O_2$ were identified, then might it become reasonable to request tens or hundreds of hours of integration time to try to observe a surface biosignature. A surface biosignature could also be observed serendipitously if exploratory long integration time observations were made of a target of major interest (in the style of the famous Hubble Deep Field).

It will be important to follow up this investigation by testing detectability of various other types of pigments and surface biosignatures, and to test their possible spectral degeneracies (especially with mineral mimics), such as done with iron oxide and iron hydroxide in Borges et al. (2024). There exists a diverse array of inorganic minerals and aerosols which can mimic pigment absorption features if combined together in the right mixture. For example, the spectral reflectance of widespread snow or gypsum may be able to somewhat mimic the shape of the Qy feature analyzed herein (Schwieterman *et al.* 2015). It may be impossible to test all combinations of all aerosols and minerals, but at least the likely combinations should be tested to identify important degeneracies. In this study on an aquaplanet, mineral signatures are unlikely to be prominent compared to pigments, but water absorbance in the NIR as well as atmospheric methane could overlap with the Qy (Bchl *b*) feature, as we explored. It may be possible to infer a totally water surface by the Berdyugina and Kuhn (2019) method of mapping surface features (if instrument capabilities allow this to be carried out in practice). If an aquaplanet scenario was suspected, it could lend support for ruling out mineral mimics.

Additional studies should be carried out such as that one and the present work, to carefully test larger instrument and biosignature parameter spaces. Another interesting next step to the present work would be to experiment with Bayesian retrievals of a diversity of surface biosignatures features on synthetic observations, such as done for the VRE by Gomez Barrientos *et al.* (2023).

## 5. Conclusions

The method used in the present work is easily generalizable to test the detectability of many different types of surface biosignatures on different planets. Carrying out surface biosignatures detectability simulations with a

3D GCM in combination with PSG allows high levels of accuracy in modeling, at the cost of increased computation time. Using a 3D GCM and parameterizing organismal abundance to be a function of climatological parameters lends accuracy and self-consistency to observation simulations that would be difficult to estimate a priori using only 1D models, and lends insight into potential observation strategies and problems. However, the magnitude to which this enhanced realism affects results is unknown, and 1D climate models may suffice for some purposes. The results presented herein broadly agree with those of Borges *et al.* (2024), which is encouraging. The integration times and distance limits that they present are, broadly speaking, the same order of magnitude as those we calculate here. However, it is difficult to compare the results precisely, due to differences in methodology.

The results of the present work suggest that a NIR surface biosignature may be detectable on Proxima b and other close-by exoplanets using future observatories (if biology is present), and suggest that such a biosignature may be detectable within 15pc. This work suggests that surface biosignatures should be readily detectable on Proterozoic-like exo-Earths if the organisms are highly abundant (with high spectral contrast) and the planetary distances not too large. It is more challenging to detect surface biosignatures on *Archean*-like exo-Earths because of their more opaque atmospheres. However, with moderately opaque atmospheres (such as *Archean Low*, or the haze simulations), surface biosignature detectability is not impeded significantly. Despite the uncertain nature of some of the assumptions made in this work (particularly the parameters of the future space observatories), we believe the results presented herein to be instructive order of magnitude detectability calculations. Generally speaking, for clear atmospheres and high organismal abundances, the Qy feature would be detectable on Proxima b in a matter of hours, and in some tens of hours at moderate abundances. The threshold for detection in under 1000 hours is a few percent global area coverage (or linear dilution from a pure organism sample). Haze and high methane concentrations make detection harder, but not impossible. Future work should carry out surface biosignature detectability calculations for other planets.

**Acknowledgements**

Thanks to Michael Way of NASA GISS for rerunning the Proxima b ROCKE-3D experiments to output the needed cloud diagnostics for PSG. C.O.M. thanks Giada Arney for helpful conversations which contributed to this work. C.O.M. would like to thank Jacqueline Austermann and Spahr Webb of Columbia University and Lamont Doherty Earth Observatory for their helpful comments during the development stages of the present work. C.O.M. dedicates this work to his dad, who always sparked his curiosity and joy for exploring the unknown, and thanks him for his support throughout the writing process of the present work.

C.O.M. acknowledges financial support from NASA OSTEM Internship Office acquired through the mentorship of G.L.V., and thanks his current faculty advisor Cheng Li for his flexibility and financial support during the latter half of the writing of the present document. N.Y.K. was supported by the NASA Earth and Planetary Science Division Research Programs, through the Internal Scientist Funding Model (ISFM) work package ROCKE-3D at The Goddard Institute for Space Studies. N.Y.K. and M.N.P. performed this work as part of NASA's Virtual Planetary Laboratory, supported by the National Aeronautics and Space Administration through the NASA Astrobiology Institute under solicitation NNH12ZDA002C and Cooperative Agreement Number NNA13AA93A, and by the NASA Astrobiology Program under grant 80NSSC18K0829 as part of the Nexus for Exoplanet System Science (NExSS) research coordination network. G.L.V. and V.K. acknowledge support from the GSFC Sellers Exoplanet Environments Collaboration (SEEC), which is funded in part by the NASA Planetary Science Divisions Internal Scientist Funding Model.

**Supporting Data**
The simulated spectra plotted or used to calculate Figures 5, 7, and 8 are available in the data.tar.gz package. The DG19 GCM states used in this work are available at: https://portal.nccs.nasa.gov/GISS_modelE/ROCKE-3D/publication-supplements/DelGenio2019Astrobio19-Habitable_Climate_Scenarios_for_Proxima_Centauri_B/. The spectra and noise simulation are generated using the publicly available PSG model (https://psg.gsfc.nasa.gov/). The scripts used to ingest the ROCKE-3D GCM data into PSG via the GlobES interface are available at: https://github.com/nasapsg/globes. The PSG instrument configuration parameters for LUVOIR are available at https://psg.gsfc.nasa.gov/data/instruments/LUVOIR_A-NIR.txt, https://psg.gsfc.nasa.gov/data/instruments/LUVOIR_A-VIS.txt, https://psg.gsfc.nasa.gov/data/instruments/LUVOIR_B-NIR.txt, https://psg.gsfc.nasa.gov/data/instruments/LUVOIR_B-VIS.txt.